\documentclass[a4paper, amsfonts, amssymb, amsmath, reprint, showkeys, nofootinbib, twoside]{revtex4-1}
\usepackage[english]{babel}
\usepackage[utf8]{inputenc}
\usepackage[colorinlistoftodos, color=green!40, prependcaption]{todonotes}
\usepackage[pdftex, pdftitle={Article}, pdfauthor={Author}]{hyperref} 
\begin{document}
\title{Emergent helicity in 
free-standing semiflexible, charged polymers}

\author{Debarshi Mitra and Apratim Chatterji}
    \email[Correspondence email address: ]{apratim@iiserpune.ac.in}
    \affiliation{Dept. of Physics, IISER-Pune, India-411008.}

\date{\today} 

\begin{abstract}
Helical motifs are ubiquitious in macromolecular systems. The mechanism of spontaneous emergence of  helicity is unknown,
especially in cases where  torsional interactions are absent. Emergence of helical order needs coordinated 
organization over long distances in  polymeric macromolecules.
We establish a very generic mechanism to obtain spontaneous helicity by inducing screened Coulomb interactions 
between monomers in a semiflexible heteropolymer.  Due to changes in solvent conditions, different segments (monomers) of 
a  polymeric chain can get locally charged  with  charges of differing polarities and magnitudes along the chain contour. 
This in turn  leads to spontaneous emergence of   transient helical structures along  the chain contour for a wide range 
of Debye-lengths.
We have avoided using torsional potentials to obtain helical structures and rely 
only on radially symmetric interactions.  Lastly, transient helices can be made long-lived 
when they are subjected to geometric confinement, which can emerge in experimental realizations through a variety of conditions.
\end{abstract}


\maketitle

\section{Introduction}

The helical motif is a recurrent feature of several macro-molecular systems and is found in disparate systems in nature. 
These systems may be as varied as the chromosome of bacterial cells \cite{kleckner_helix} and the secondary structure of proteins or actin filaments
or in collagen molecules \cite{actin,collagen,Banavar2004,ecoli_nucleoid}. Thus, the helical structure is  ubiquitous in the macro-molecular world. 
This has spurred a number of studies that investigate their emergence in several natural systems \cite{Maritan2000,Pokroy2009,Sabeur2008,Li2018,Vega2019,Snir2005,Banavar2004,Gerbode2012,Forterre2011,Biswas2021,Chaudhuri2012}. 
Furthermore,   helical motifs in  
synthetic molecules have been explored for their key technological ramifications \cite{Srivastava2010} such as the synthesis of NEMs devices \cite{Nakano2001,Zhong2008,Jiang2013,Singh2014,synthetic_helix,kotov}, piezoelectric devices \cite{Kong2003} and other optical materials \cite{Stratford2014,Lei2018,feng2017assembly}. 
For these applications, helical molecules are typically designed using techniques such as vapour deposition which rely on specific 
interactions between the  constituent monomers. Alternatively, helical templates are also often employed \cite{Wang2013,Wang2008,Zhang2002,Zhang2003,Liu2011,Rapaport2002}. 

It would be of interest to unearth mechanisms by which one may induce chains to spontaneously self-organize 
to obtain helical structures at the $nm$ to $\mu m$ length scales. If these mechanisms are generic 
in nature and independent of the chemical details of the monomers, then one can have an alternative strategy 
of designing synthetic helical macro-molecules in the laboratory. 
More significantly, such studies would also shed light on the mechanisms governing the formation of helical structures 
in nature, both inside and outside the living cell.

Previous reports have demonstrated the formation of extremely short-lived helical polymers in bad solvents undergoing collapse \cite{Sabeur2008}.
The packing of tubular filaments in confined cylinders also lead to helix formation
for  particular  ratios of pitch and radius \cite{Maritan2000}. Yet, another study shows that the
ground state of a self-attracting chain shows a variety of o
structural motifs including the helix, depending on  the nature of
the stiffness present in the chain (energetic or entropic) 
\cite{krbi2019,Snir2005}. In our previous work we showed that a bead-spring model of 
a semiflexible polymer chain with monomers all having the same value and  polarity of charges and long range Coulomb interaction
between  monomers attains a transient helical conformation, when repulsive  interactions are switched 
on between the monomers \cite{Mitra2020}. The monomers may become charged because of a change in the nature of the solvent, or if the 
ionic concentration of the solvent is suddenly changed or due to other changes in its environment. We obtained transient 
helices even when we replaced Coulomb interactions by long range repulsive power-law interactions.

Here we introduce a more generic model and establish that a semiflexible polymer chain with  charged beads,  each carrying a  
different value of  charge and polarity, can also attain a transient helical conformation even due to the short range 
screened (spherically symmetric) Coulomb interactions between the monomers. This scenario is likely to occur in many natural systems. Additionally, we show that asymmetry in the distribution of charges along the 
polymer chain plays a pivotal role in helix formation. A chain with an approximately equal number of positive and negative 
charges does not form helices. 
The helix formation takes place only for a certain range of parameters of the polymer, such as the persistence length, 
the spring constant, the Debye length etc. We further show that the helices formed by these screened electrostatic interactions 
can be made long-lived under geometric confinement, in a cuboid of appropriate dimensions. 

In the case of bacterial nucleoids, 
this geometric confinement can be imposed by the bacterial cell leading to stable helical structures of nucleoids. 
These beads can be molecular units of a heteropolymer, or groups of amino acid sequences with length scale of the size 
of the beads, or even larger units of molecules such 
as nucleosome complex in the DNA dressed by suitable (charged) proteins, or even polymeric chains where the monomeric units are micron sized 
colloids. In this case, the colloidal surface can be dressed by microgel particles which have been cross-linked together as shown by \cite{Biswas2017,Biswas2021}.  The colloidal particles then form a polymeric chain held together by cross-links with tunable semiflexibility (persistence length). 
If one uses colloids that are chemically different to each other, to form the chain, a change in the ionic concentration in the solution can result in a different
value of charge on each colloid. 



\section{Model}
We use { Brownian dynamics simulations} to study helix formation in a single semiflexible, 
bead-spring model of a polymer chain, where the monomers have different values of charges 
along the chain contour as determined by a probability distribution. The center of each monomer is
connected by harmonic springs of spring constant $\kappa$ and of equilibrium length $a$ to the 
center of two neighbouring monomers. { We choose $a$ to be the unit of length for our simulations.} 
Furthermore, there is a bending energy cost per every triplet of 
monomers along the chain contour {to incorporate effects of semiflexibility}. 
{The energy cost of having bends along the chain, is realized} by the bending potential  $V(\theta)=\epsilon_B \cos(\theta)$, where
$\theta$ is the angle between the bond vectors ($-\mathbf{r}_i, \mathbf{r}_{i+1}$). The vector 
$\mathbf{r}_i$ is the vector joining monomer $i-1$ to its neighbouring monomer $i$ along the chain contour.
The quantity $\epsilon_B$ can also be directly related to the persistence length $\ell_p$,
\begin{equation}
    \ell_p \approx a \epsilon_B/k_BT
\end{equation}

for small deviations of $\theta$ from $\pi$, refer appendix  of \cite{Mitra2020}. In our simulations, energy 
is measured in units of the thermal energy $k_BT$. The excluded volume of the monomers are modelled by the WCA 
(Weeks, Chandler, Andersen) potential  and all the monomers have diameters $\sigma=0.7a$. 


Additionally, any two monomers $i$ and $j$ situated at a distance $r_{ij}$ from each other, 
with charges $q_i$  and $q_j$  interact with each other through the screened Coulomb interaction. 
\begin{equation}
V(r_{ij}) =  \frac{ q_i q_j}{4 \pi \epsilon r_{ij}} \times  \exp^{(-r_{ij}/\lambda_D)} ...\forall r_{ij} < r_c 
\end{equation}
 where the  Debye length $\lambda_D$ is a parameter and sets the length scale over which Coulomb interactions get screened out.
The permittivity $\epsilon$ of the background medium within which the chain is suspended, can be expressed in terms of 
the permittivity of vacuum $\epsilon_0$ and the  relative permittivity $\epsilon_r$.
The strength of interactions is determined by the value of $q_i $ and $q_j $, but for 
our calculation we will express charge in units of $Z_0$, such that the interaction between any two charged monomers can be expressed as:
\begin{equation}
V(r_{ij}) =  Z_i Z_j/r_{ij} \times  \exp^{(-r_{ij}/\lambda_D)} ...\forall r_{ij} < r_c 
\end{equation}
where, $Z_i$ and $Z_j$ denote the value of charges and is expressed in units of $Z_0$, on the interacting monomers in simulation units. 
In the subsequent section, we elaborate on how the values of $Z$ can be related to charge in real units. \\

We assign charges to the monomers in the chain such that there is stochasticity in the distribution of charges along the polymer chain.
The charge $Z_i$ of each monomer $i$ in simulations will be allocated in terms 
of the asymmetry parameter $\psi$ and $Z_0$ such that $Z_i = Z_0 \times (r - \psi)$, where $r$ is a random number whose
values are chosen from a uniform distribution of random numbers between $0$ and $1$.
We vary $\psi$ between 
$0$ and $1$ to estimate the net charge of the chain. A value of $\psi=1$  
implies that all monomers have random values of charges between $Z_1=0$ and $Z_2=-Z_0$ s.u.
 A value of $\psi=0$  
implies that all monomers have random values of charges between $Z_1=0$ and $Z_2=Z_0$ s.u.
Correspondingly, for $\psi=0.5$, the charge on the monomers can take random values between $Z_1=-Z_0/2$ to  
$Z_2=Z_0/2$ s.u. Thus, in general, the charge on a monomer $i$ can vary between $Z_1$ and $Z_2$.

We will establish later in the manuscript that if $\psi=0.5$ such that there are statistically  equal number of positive and negative charges, then we do not obtain helices.
We need to introduce asymmetry in the distribution of charges using $\psi$,  such that $Z_1 \neq Z_2$.
The degree of asymmetry in the distribution of charges along the polymer chain 
is denoted by  {the quantity } $\psi$. 
Thus, $\psi=0.1$  corresponds to  a chain that statistically has $90\%$ 
of the monomers having a positive charge.

 For our  Brownian dynamics simulations where the friction constant is $\mathbf{\zeta}$, 
and the unit of time $\tau$ is set by $\tau=a^2 \zeta/k_BT$, the time taken  for an isolated monomer
to diffuse a distance of $a$..
If we set $\zeta=1$ such that $\tau=1$ with $k_BT=1$ and $a=1$ (units of energy and length, respectively, 
the over-damped stochastic Brownian  dynamics equation is integrated with time step $dt =0.0001 \tau$.

\subsection{Relevance to experimental systems}
We now establish a correspondence between the magnitudes of charges used in the simulation (expressed in s.u.), 
and the magnitudes of charges in real units. If two monomers having unit charges 
in the simulations, are at a unit distance away from each other then they interact with Coulomb potential energy of strength $k_BT$. Let $q_o$ be the charge in real units, corresponding to $Z=1 s.u$, where $Z$ denotes the charge on a monomer in the simulations.

In real units two charges ($q_0$) of identical charges have potential energy given by
\begin{equation}
\frac{(q_0)^2}{4\pi\epsilon a} =\frac{Z^2}{a}=k_BT
\end{equation}
Thus at room temperature of  $T=300K$, the value of $q_o$
in water ($\epsilon_r=80$) placed at unit distance ($a$) is given by,

\begin{equation}
q_o=\sqrt{4\pi\epsilon a k_BT} =\sqrt{37 \times 10^{-30}a} \,  C.
\end{equation}
where the charge $q_0$ is measured in Coulomb (C).
If we choose $a=1nm$, then $q_o=1.9 \times 10^{-19}$C, or $1.2e$, where $e$ denotes the electronic charge. 
If $a=10nm$ then we obtain $q_o\approx 4e$.
If $a=100 nm$ then, $q_o\approx 12 e$ 
and if $a=1 \mu m$ then  $q_o \approx 40e$. 
In this study we often use $Z_0 \sim 10 s.u$ which corresponds to a charge of $40 e$ for a
monomer with size $10$ nm. But that is the maximum charge on a monomer for $\psi=1$. For lower values of $\psi$, the maximum 
charge that a monomer can take will be lower. 
Thus, the charges used in this study may feasibly occur in many macro-molecular systems.



 \begin{figure*}[!hbt]
\includegraphics[width=0.8\columnwidth,angle=0]{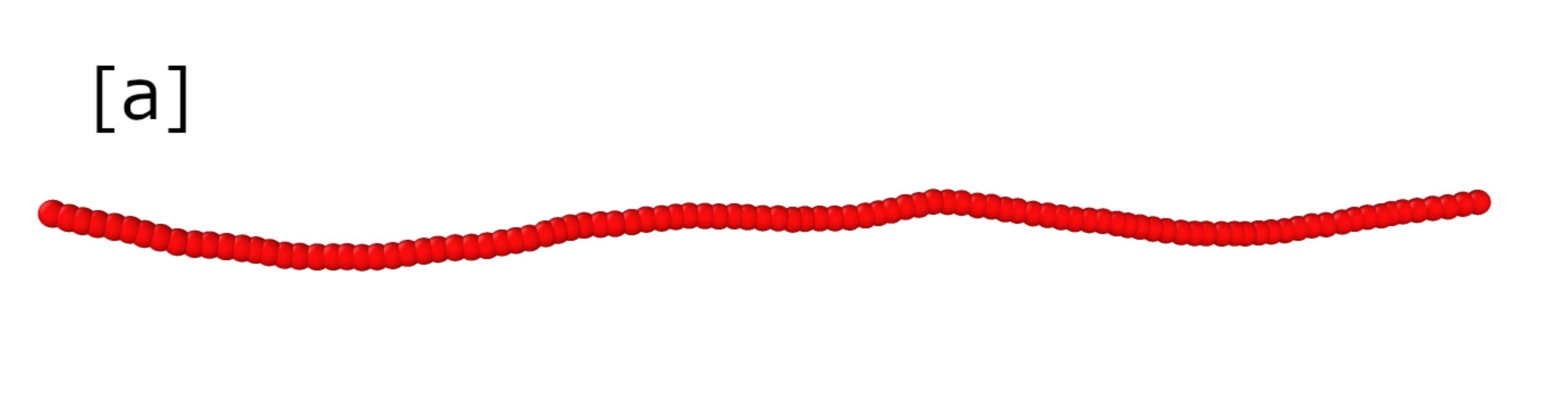}
\includegraphics[width=0.8\columnwidth,angle=0]{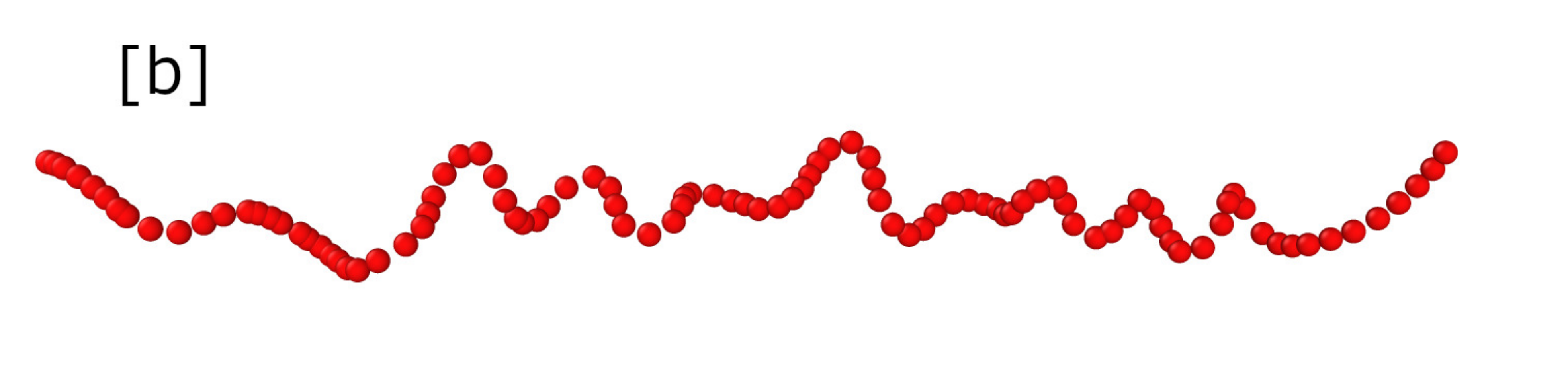}
\includegraphics[width=0.8\columnwidth,angle=0]{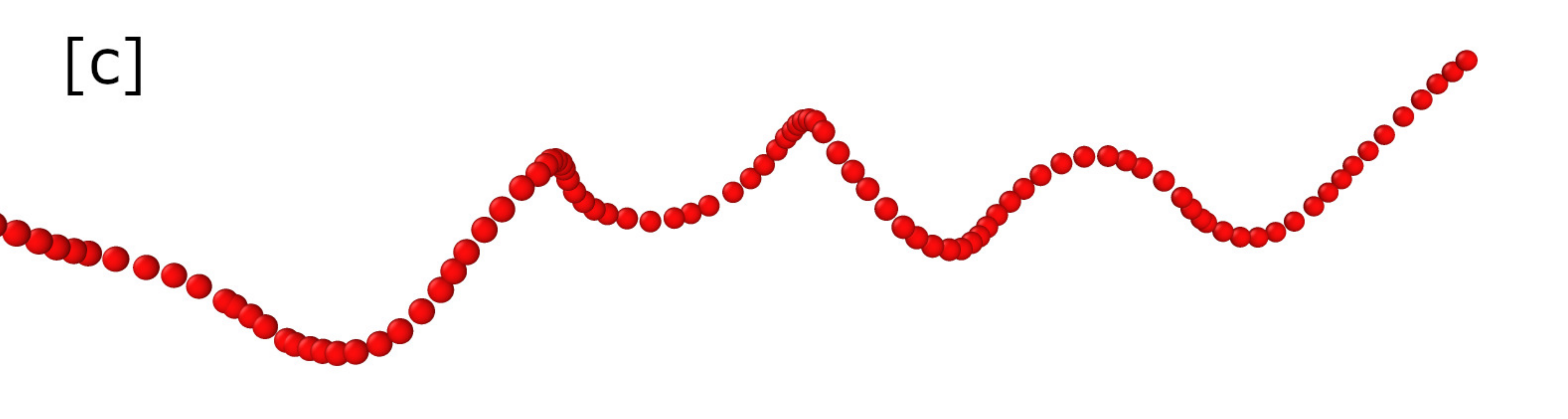}\\
\vskip0.7cm
\includegraphics[width=0.6\columnwidth]{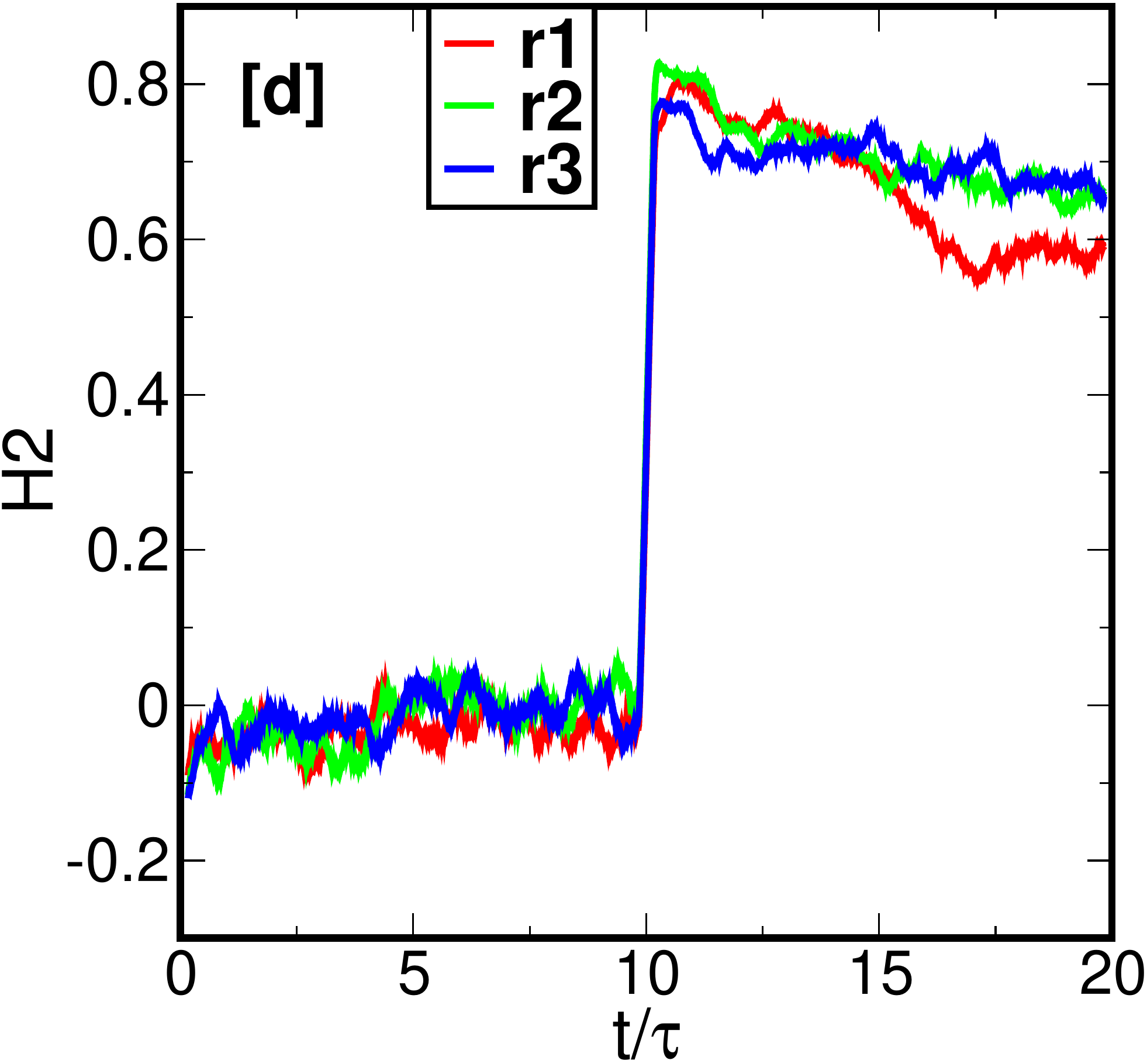}
\includegraphics[width=0.6\columnwidth]{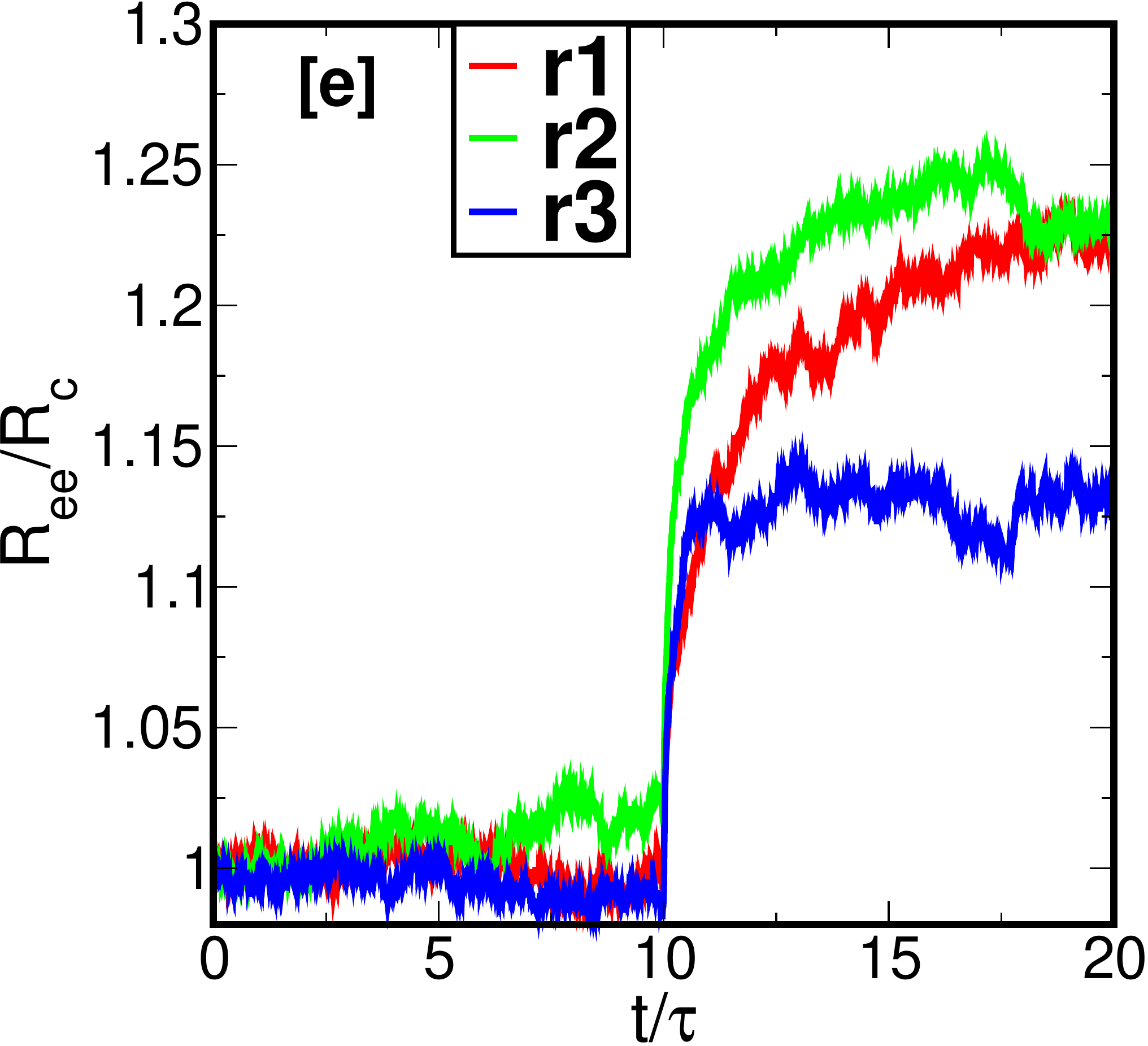} 
\caption{\label{fig1}
{\bf Helix formation:}
The snapshots shows various stages of the helix formation in a bead-spring model of a
semi-flexible polymeric chain with $100$ monomers, with contour length $R_c=99a$. Initially, the uncharged polymer is in equilibrium 
with a thermal bath : representative snapshot shown in [a] at time $t=3 \tau$. Snapshot [b] shows a representative 
conformation  of the chain soon after  the monomers acquires charge at time $t=10 \tau$, where we see formation
of helices along the chain contour. At (c) $t=20\tau$, we see that the helix gradually ``unwinds" and relaxes to 
a helical conformation having a larger pitch. The charge on each monomer has a random value,  between  $-4.0$ to
$36$ s.u. (simulation unit). This corresponds to a asymmetry parameter of $\psi=0.1$ (refer to the Methods section for the definition of $\psi$)  and $Z_0=40$ s.u.
The bending energy cost $\epsilon_B=200 k_BT$ (corresponding to a persistence length of $200a$ and spring constant 
$\kappa=100 k_BT/a^2$, and the Debye  length $\lambda_d=5a$.  
In (d) we show the helical order parameter $H2$ (running averaged over $0.33 \tau$) as a function of 
simulation time $t/\tau$, for $3$ independent runs. We see a sharp rise in $H2$ at time $t=10 \tau$, which
decays thereafter, indicating transient helix formation. Subfigure (e) shows the end to end distance 
$R_{ee}/R_c$ versus $t/\tau$. 
}
\end{figure*}

\begin{figure*}[!hbt]
\hskip0.1cm
\includegraphics[width=0.48\columnwidth]{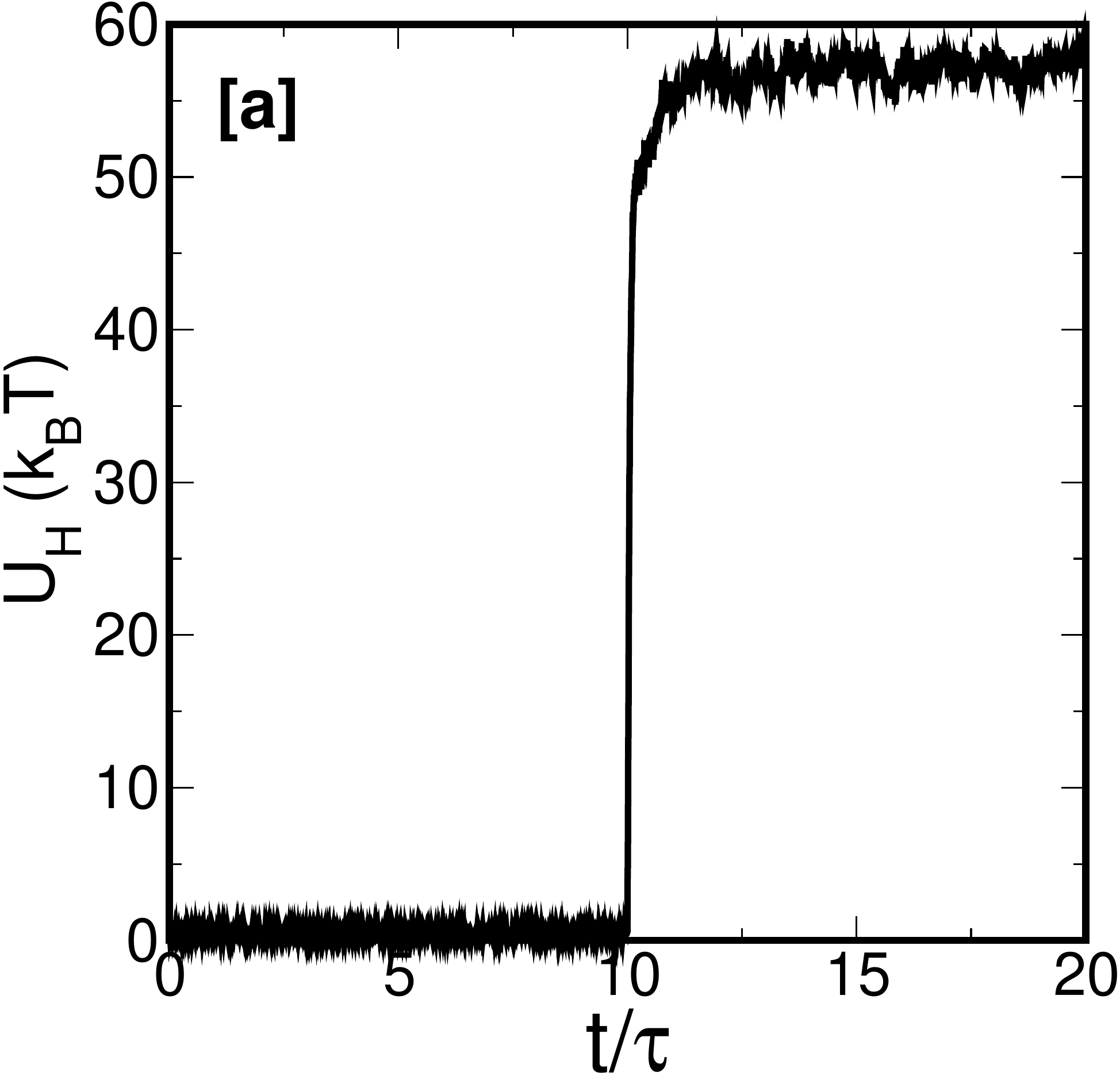}
\includegraphics[width=0.48\columnwidth]{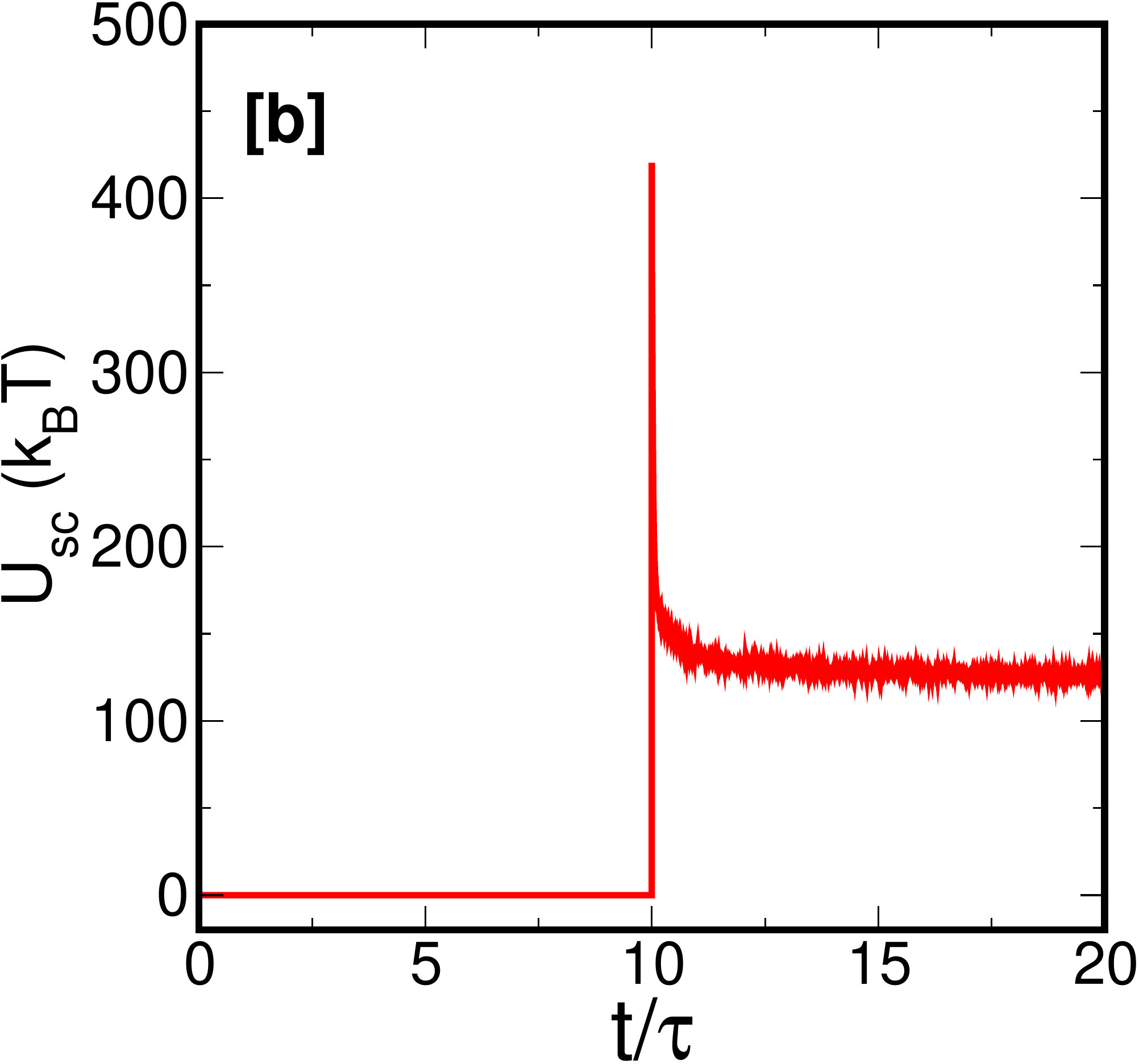}
\includegraphics[width=0.48\columnwidth]{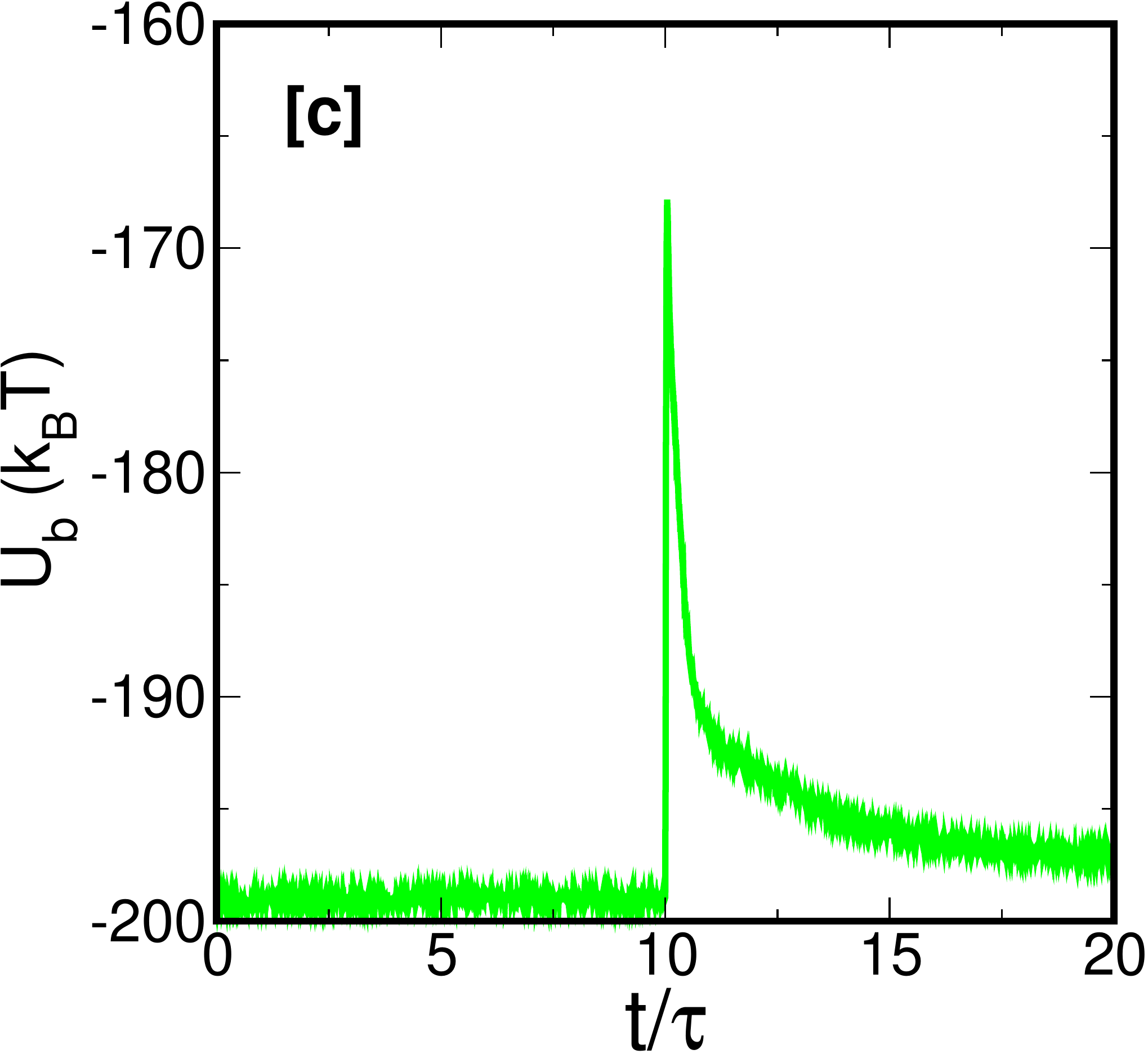}
\includegraphics[width=0.46\columnwidth]{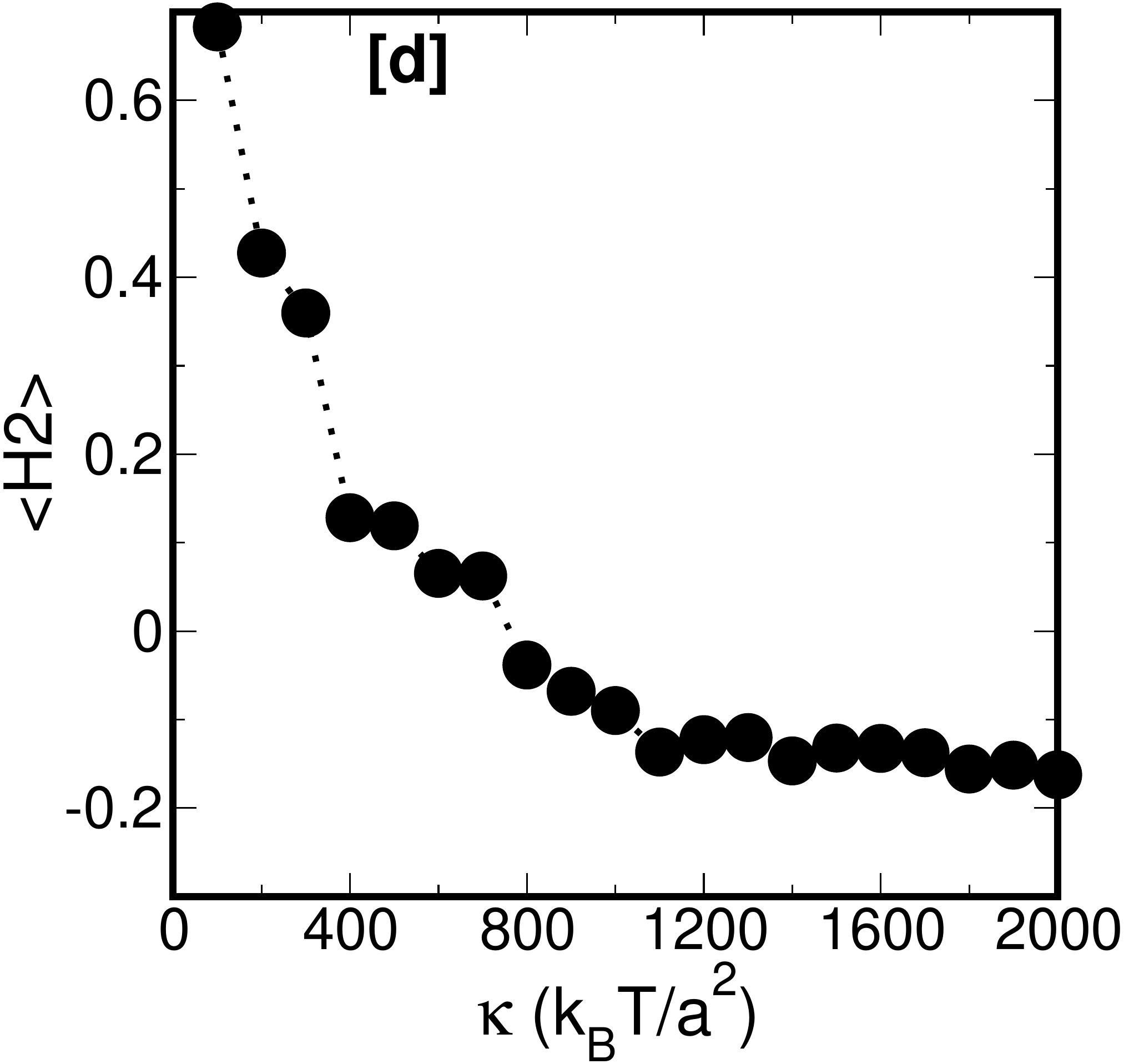}
\caption{\label{fig2}
{\bf Mechanism of Helix formation: }
In (a), (b) and (c) we show the harmonic spring energy ($U_H$) , the screened coulomb energy ($U_{sc}$) and the bending energy ($U_b$) per monomer. We see that just after the screened Coulomb interaction is switched on, the screened coulomb energy and the bending energies peak, and subsequently relax leading to the ``unwinding" of the helix. The springs in the polymer chains however continue to remain stretched due to the screened Coulomb interactions, as shown in (a). The sudden switching on of screened Coulomb interactions at $t=10\tau$, leads to the formation of sharp angular conformations. This can vbe inferred by noting that $U_b$ shows a peak at this point,in the plot given in (b). Subsequently, the transient helix formation takes place by the gradual relaxation of these sharp bends. Furthermore, if the springs are made too stiff, these sharp bends are not formed since a high value of $\kappa$ prevents the monomer-bonds from stretching. Thus, in (d) we show that $\langle H2 \rangle$ shows a monotonic decrease with increase in $\kappa$. We calculate the time averaged $\langle H2 \rangle$ over $10 \tau$ to $20 \tau$. The values of other parameters are same as that of the data presented in Fig.\ref{fig1}.}
\end{figure*}

\section{Results} 
\subsection{Helix Formation}
We observe spontaneous emergence of helices if a uncharged semiflexible polymeric chain acquired charges at time 
$t=T_0$ due to change in external conditions, e.g. change in pH of the solvent in which it is immersed.
The helices formed are transient but may be stabilized by other factors such as hydrogen bond and geometric 
confinement, as we show later. We establish that the stochasticity in the value or sign of the charges on the 
monomers along the  chain contour do not affect helix formation as long as there is sufficient asymmetry 
(as determined by $\psi$) in the charge distribution. We have tried out different initial conditions, 
such that monomers $i-1, i, i+1, i+2$ have different value of charges in different runs for a particular
value of $\psi$ (refer to the Methods section for the definition of $\psi$)  and monitor the helicity developed as we switch on charges on the monomers.

We measure helicity using the order parameters $H2$ which is defined as 
 \begin{equation}
 H2=\frac{1}{N-3}(\sum_{i=2}^{i=N-2} \mathbf{u}_i.\mathbf{u}_{i+1})
\end{equation} 
 where $\mathbf{u}_i$ is the unit vector of $ \mathbf{U}_i =
 \mathbf{r}_i \times  \mathbf{r}_{i+1}$.
 A simple uncharged, semi-flexible polymer chain 
 shows $H2$ values $\approx 0$ (or negative values of $H2$)
as expected for a chain locally bent due to thermal
fluctuations, while calculating $H2$ over  helical segments of a polymer chain would lead to high 
positive values of $H2$ \cite{Biswas2021,Mitra2020}. 

Past studies \cite{Sabeur2008,Mitra2020} have also used another order parameter which is given by,

\begin{equation}
 H4= \frac{1}{N-2} (\sum_{i=2}^{i=N-1} \mathbf{u}_i)^2; 
\end{equation} 
A perfect helix with uniform handedness, will have vectors $u_i$ pointing
along the helix axis, and hence $H4$ will have a value of
$\approx 1$. However, if one obtains a helical polymer where
half of the chain is right-handed, and the rest of it is
left-handed, $H4$ will be zero. In this study, we do not have any control over the handedness of 
the helical structures and thus for long chains, the chain is likely to have an equal number of 
right-handed and left-handed segments. Thus, we use the parameter $H2$ to quantify
helicity, in the remainder of the manuscript. In the future, we will report how we can design left 
handed and right  handed helices, made to order.

We start our simulations with a uncharged semiflexible polymer chain in  a heat bath, allow 
the chain to reach equilibrium over $t=T_0= 10 \tau$ ($10^5$ iterations), and switch on charge 
on the monomers thereafter. We show the various stages of helix 
formation through representative snapshots give in Fig.\ref{fig1}(a-c). We also show that the helical polymer chain 
can easily be distinguished from a non-helical, uncharged, semiflexible polymer chain using $H2$, which fluctuates 
near zero till $t=10 \tau$ and then show a finite non-zero value. In Fig.\ref{fig1}d
we show $H2$ as a function of time $t$ for three independent runs. In Fig.\ref{fig1}e,  we plot the end to 
end distance $R_{ee}$, scaled by the contour length $R_C$ ($R_c=99a$) as a function of time. We see that the end to 
distances fluctuate  at values $\approx 1$ for three independent runs till $t=10\tau$, and thereafter the $R_{ee}/R_c$ 
shows a sudden increase. Thus we show that the helix formation is  independent of the initial configuration of the monomers
as well on the stochastic distribution of charges along the chain contour. 

{\bf Mechanism of helix formation: } The uncharged semiflexible  polymer chain undergoes
stochastic conformational fluctuations due to the thermal noise from $t=0$ to $t=10 \tau$.  When the electrostatic 
interactions are switched on, electrostatic forces leads to a reinforcement of the  angular conformations
with larger deviations in the angle $\theta$ from the chain axis.  This is because neighbouring 
charges repel each other, and move neighbouring monomers along the line joining their centers, and away from the chain axis. These angular conformations (with large anglular deviations) 
are penalised due to the semiflexibility. As the different chain segments along the chain contour locally {\em straighten and stretch} out from the axis to reduce bending energy costs, 
the polymer consequently attains a helical conformation. 

This can be surmised from Fig.\ref{fig2} where we plot the different energy contributions in the system, for parameters 
identical to that of Fig.\ref{fig1}. In Fig.\ref{fig2}a  we plot the energy per spring, $U_H$ versus time $t$.
We see that at $t < T_0$, $U_H$ fluctuates around $0.5 k_BT$, as expected from the equipartition theorem. 
However at $t=10\tau$, we see that $U_H$ shows a drastic increase which indicates that the bonds get stretched 
severely as a consequence of the electrostatic (e.s) interactions. This is further 
corroborated by the data presented in Fig.\ref{fig2}b where we plot the screened Coulomb energy per monomer $U_{sc}$ as a 
function of time. We see that at $t=10\tau$, $U_{sc}$ shows a sudden spike which subsequently relaxes to lower values. 

We also look at the bending energy $U_b$ per triplet of monomers, in Fig.\ref{fig2}c and we find that at $t=T_0=10\tau$,
the bending energy shows a sudden spike, indicating a sharply kinked (angular) conformation. However, such a conformation 
is penalised by the bending energy which prevents such sharp angles. Thus the helix formation occurs as a relaxation 
of transiently formed sharp angles. We emphasize again that  the helix formed is transient and eventually the helical 
conformation dissolves. Thereafter, the polymer chain adopts an extended linear configuration due to the electrostatic 
interactions. Furthermore, thermal  fluctuations play a pivotal role in the helix formation in creating the initial
small bends which get accentuated  after switching on the charge on the monomers. The transient helices do not form,
if we switch on charge at temperature $T=0$, as there are no thermal fluctuations. 
This has been explored in detail in \cite{Mitra2020}, previously.

The importance of the formation of kinks in the subsequent emergence of helicity can also be gauged by the 
increasing the simulation $\kappa$ parameter to high values, keeping $k_BT=1$.
We monitor the time averaged value of $H2$ with increasing $\kappa$ in Fig.\ref{fig2}d, and  observe that 
$\langle H2 \rangle$ decreases with increasing $\kappa$. An increase in the spring 
constant prevents the bonds from stretching as much, and thereby formation of kinks on switching on
of charges, and that in turn in turn suppresses  subsequent helix formation. 
 
\begin{figure}[!hbt]
\includegraphics[width=0.98\columnwidth]{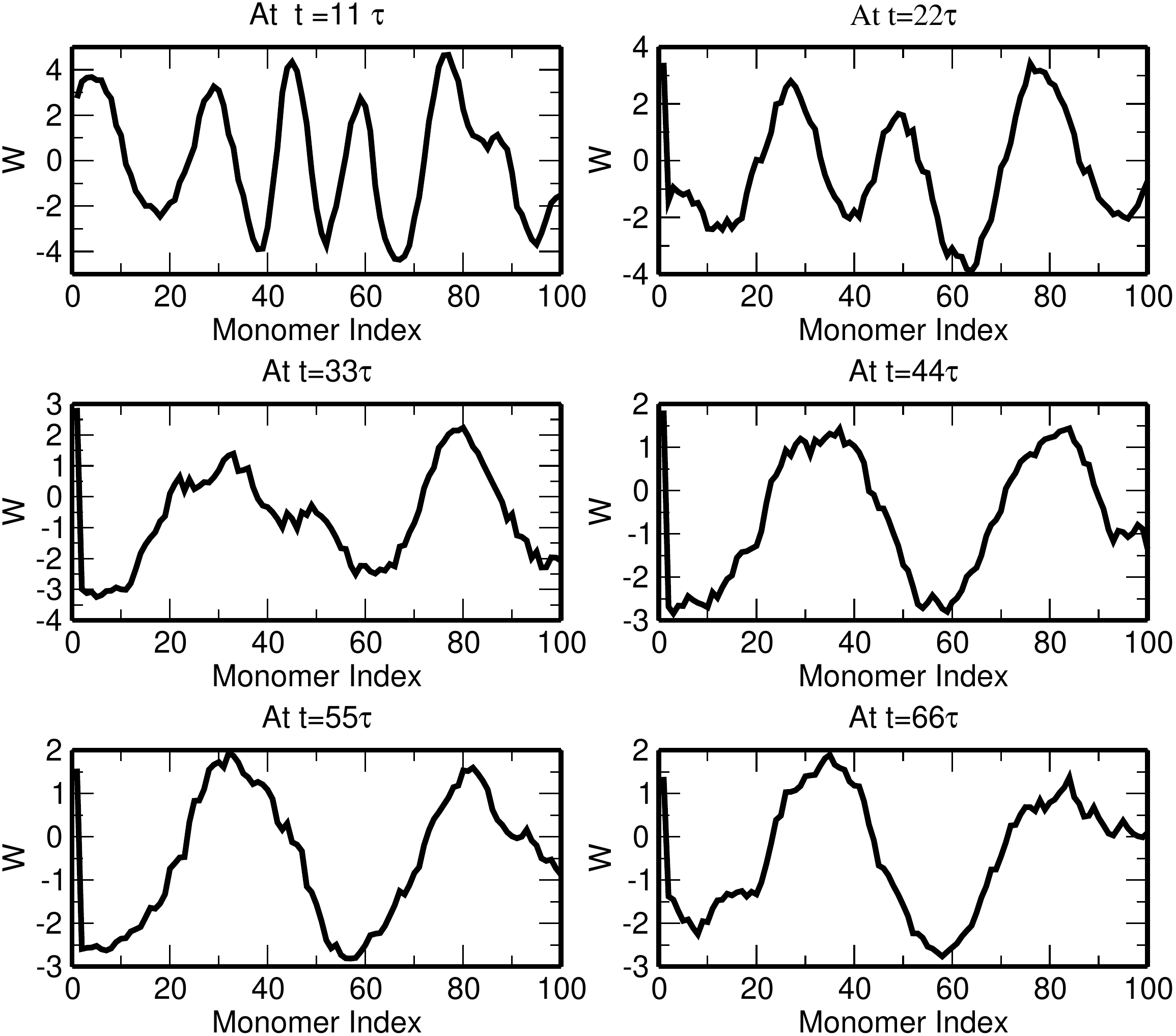}
\caption{\label{fig3}
{\bf Unwinding of the helix: Change in helical pitch} 
To quantitatively establish the unwinding of the helical polymer we look at the periodicity parameter W as a function of time. W is a scalar quantity resulting from the dot product of bond vectors along the chain contour with a vector
perpendicular to the axis of the helical polymer chain. The periodicity in W indicates the pitch of the helix. In subfigure (a) we show $W$ for different times for a chain of $100$ monomers, where we have switched on electrostatic interactions at $t=10\tau$. The other parameters are also identical to that of the data presented in Fig.1 We note that due to the gradual unwinding of the helical polymer the periodicity in W decreases.}
\end{figure}

{\bf Unwinding of the transient helices}\\
The emergent structure of helices ``unwind" with time. There are two factors that 
contribute to the unwinding: the bending energy and the screened coulomb interaction. The high 
energy cost of bending  leads to the gradual relaxation of "kinks" (sharp angles) along the chain contour and thus leads to a straightening of the chain, 
such that $\theta \approx \pi$. This results in a gradual dissolution of the helical structures.
The repulsive electrostatic interaction also leads to an ``effective" additional bending energy cost \cite{es_stiff}
and causes the polymer chain to adopt an extended configuration. Thus the helix 
formation occurs only as a kinetic pathway to an extended linear equilibrium conformation.
 
 \begin{figure*}[!hbt]
\includegraphics[width=0.48\columnwidth]{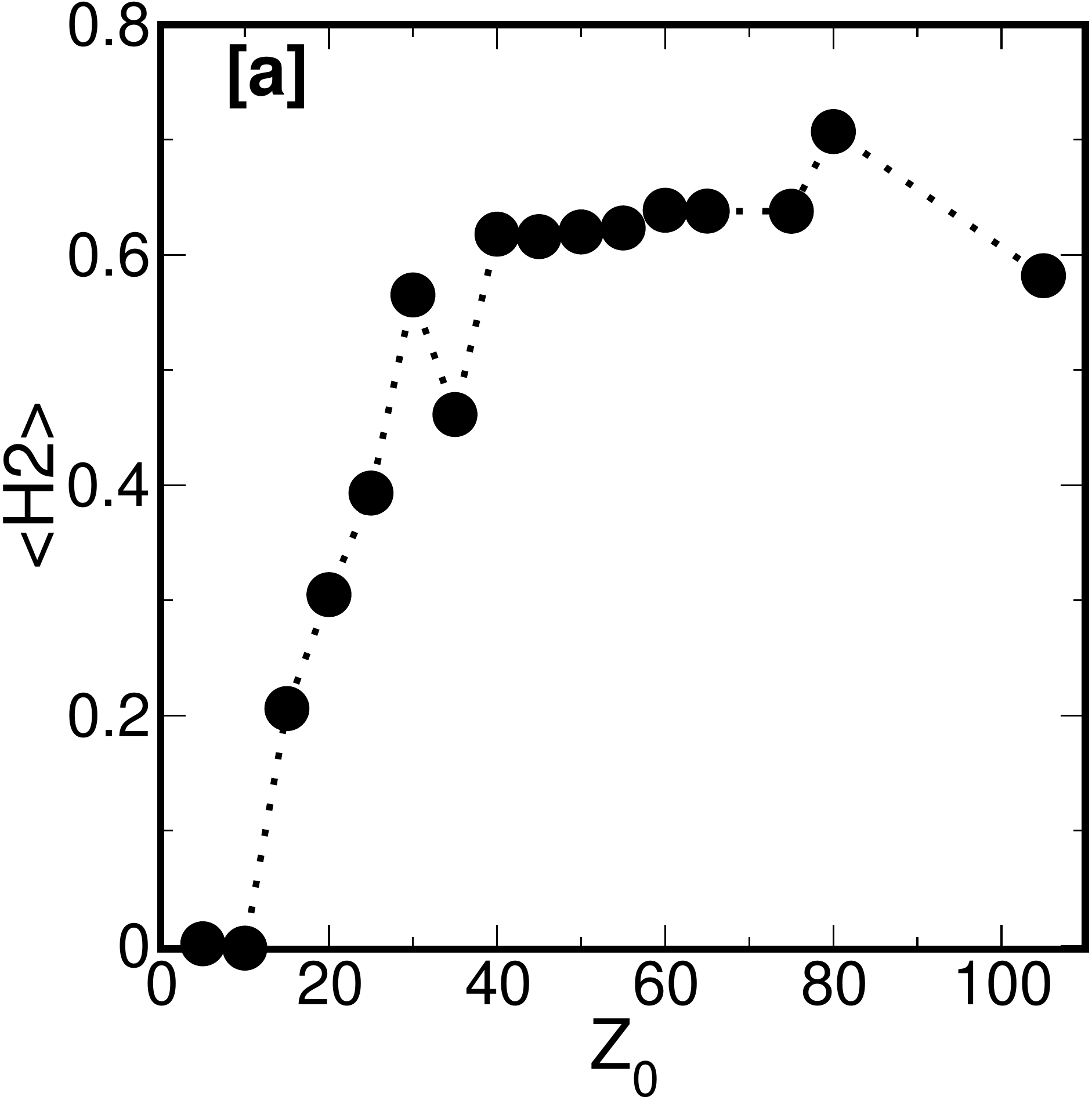}
\includegraphics[width=0.52\columnwidth]{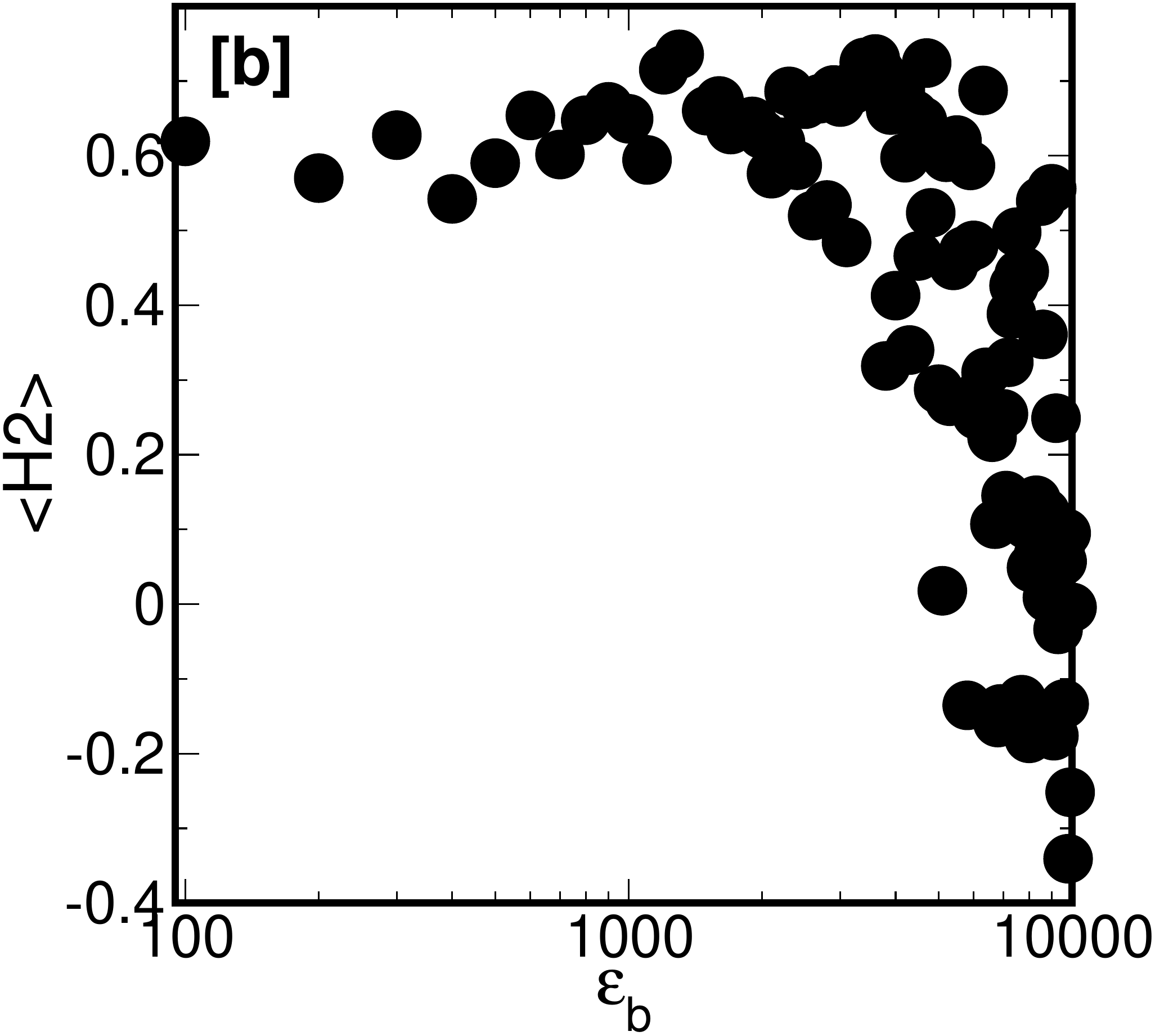}
\includegraphics[width=0.48\columnwidth]{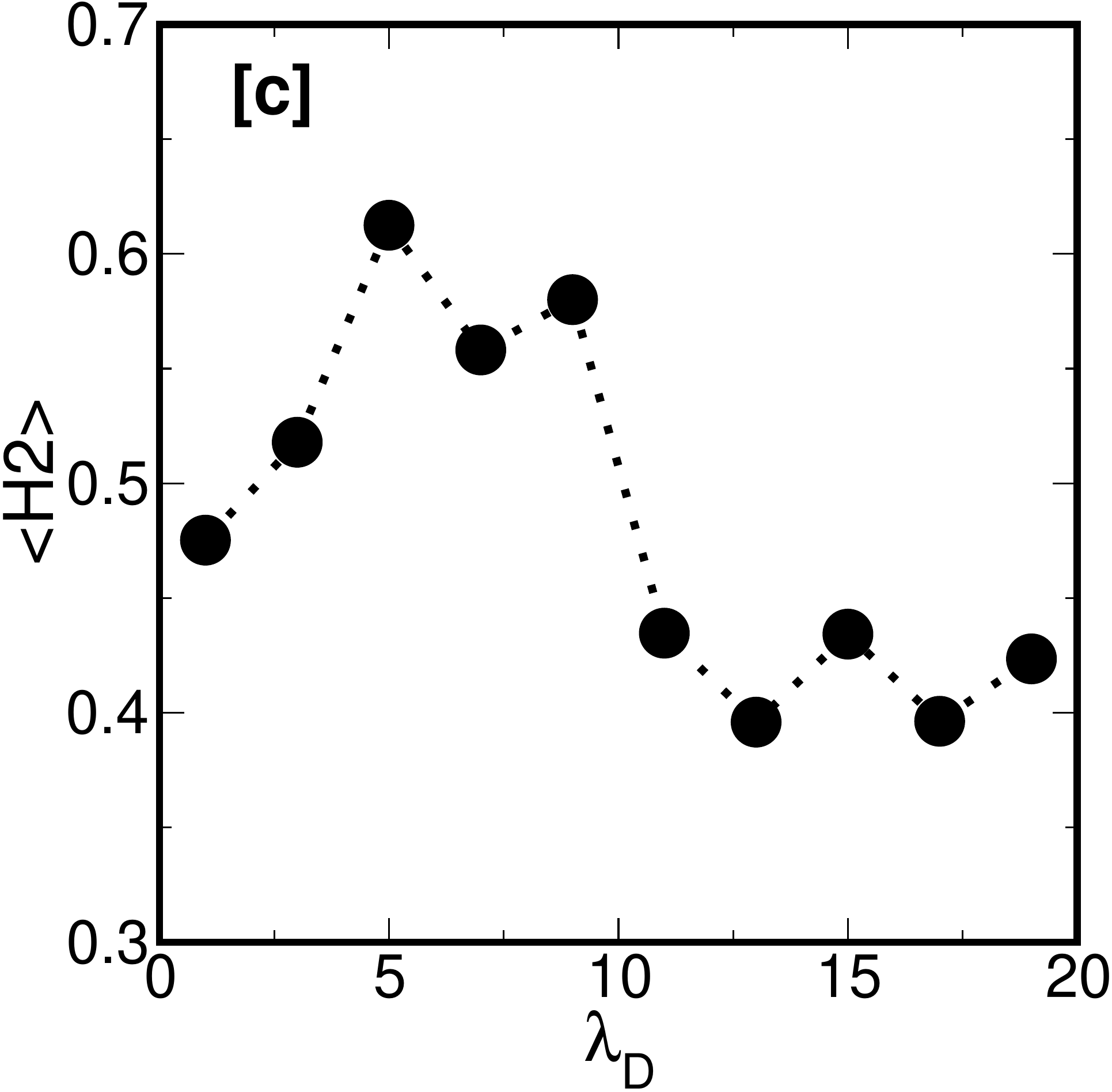}
\includegraphics[width=0.48\columnwidth]{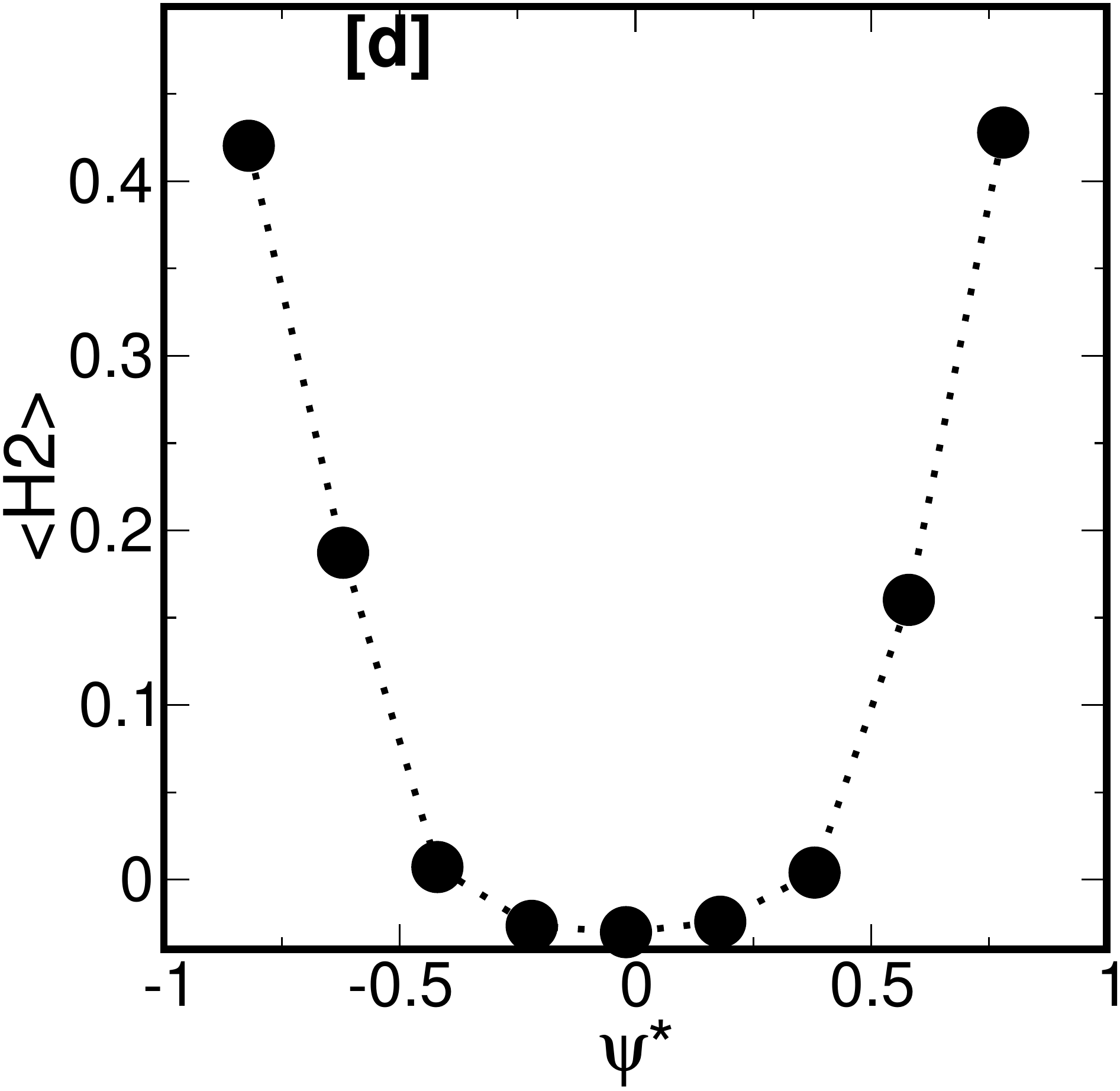}
\caption{\label{fig4}
{\bf Range of parameters conducive to helix formation:}
For this set of subfigures (a-d), we keep the parameters $Z_0$, $\epsilon_B$, $\lambda_D$, $\kappa$, and the asymmetry parameter $\psi$
identical to that in Fig.\ref{fig1}, but then  vary one parameter at a time keeping the others fixed. We plot the time 
averaged helicity order parameter $\langle H2 \rangle$ to analyze the helix formation as we modify different system parameters.
Subfigure (d) shows the dependence of $\langle H2\rangle$ on the charge asymmetry parameter $\psi*=1-2\psi$. The quantity $\psi*$ 
indicates the asymmetry in the charge distribution in the polymer chain. 
The data in (c) and (d) has been produced by averaging over $10$ independent runs.}
\end{figure*}

 \begin{figure}[!hbt]
\includegraphics[width=0.9\columnwidth]{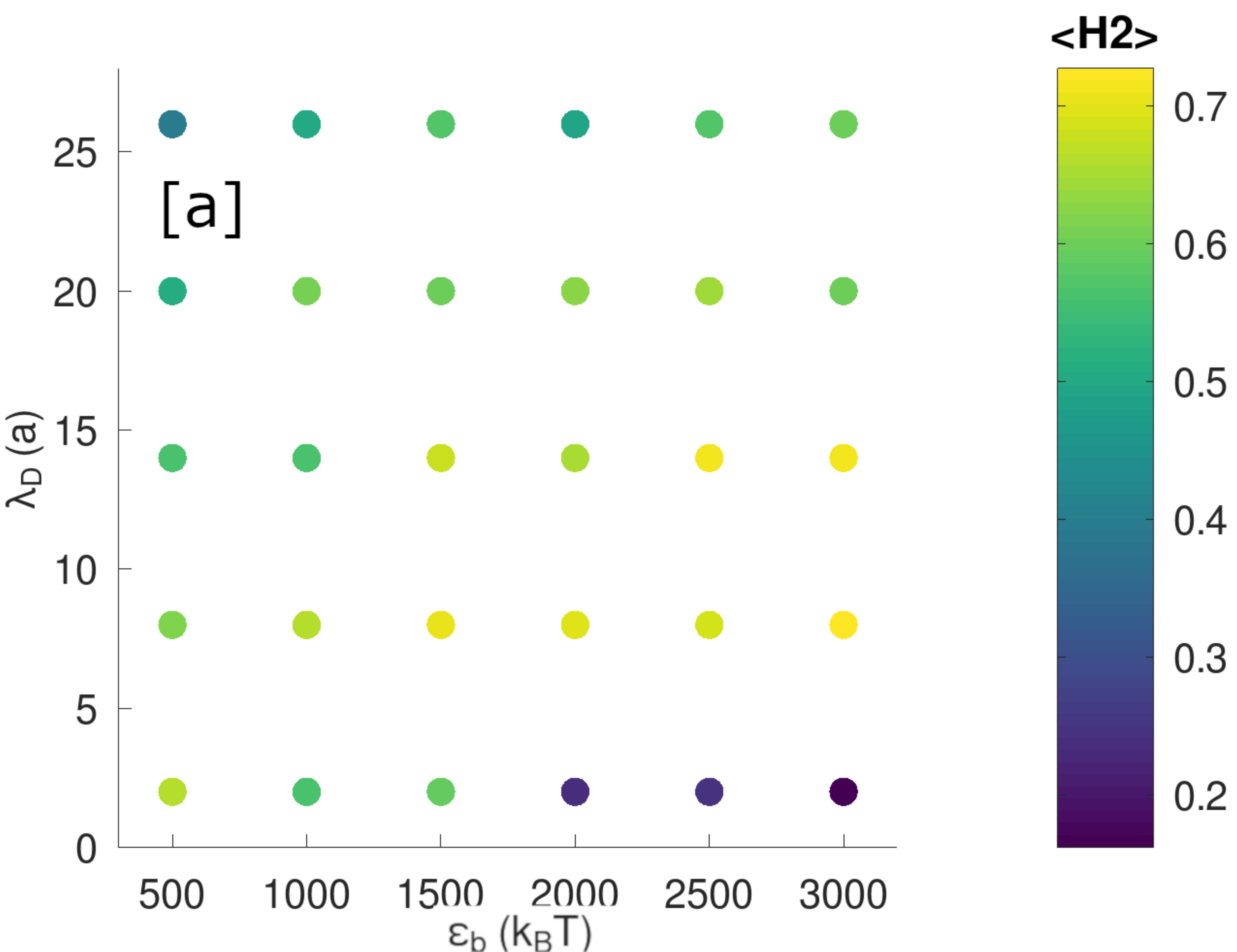}
\vskip0.2cm
\includegraphics[width=0.9\columnwidth]{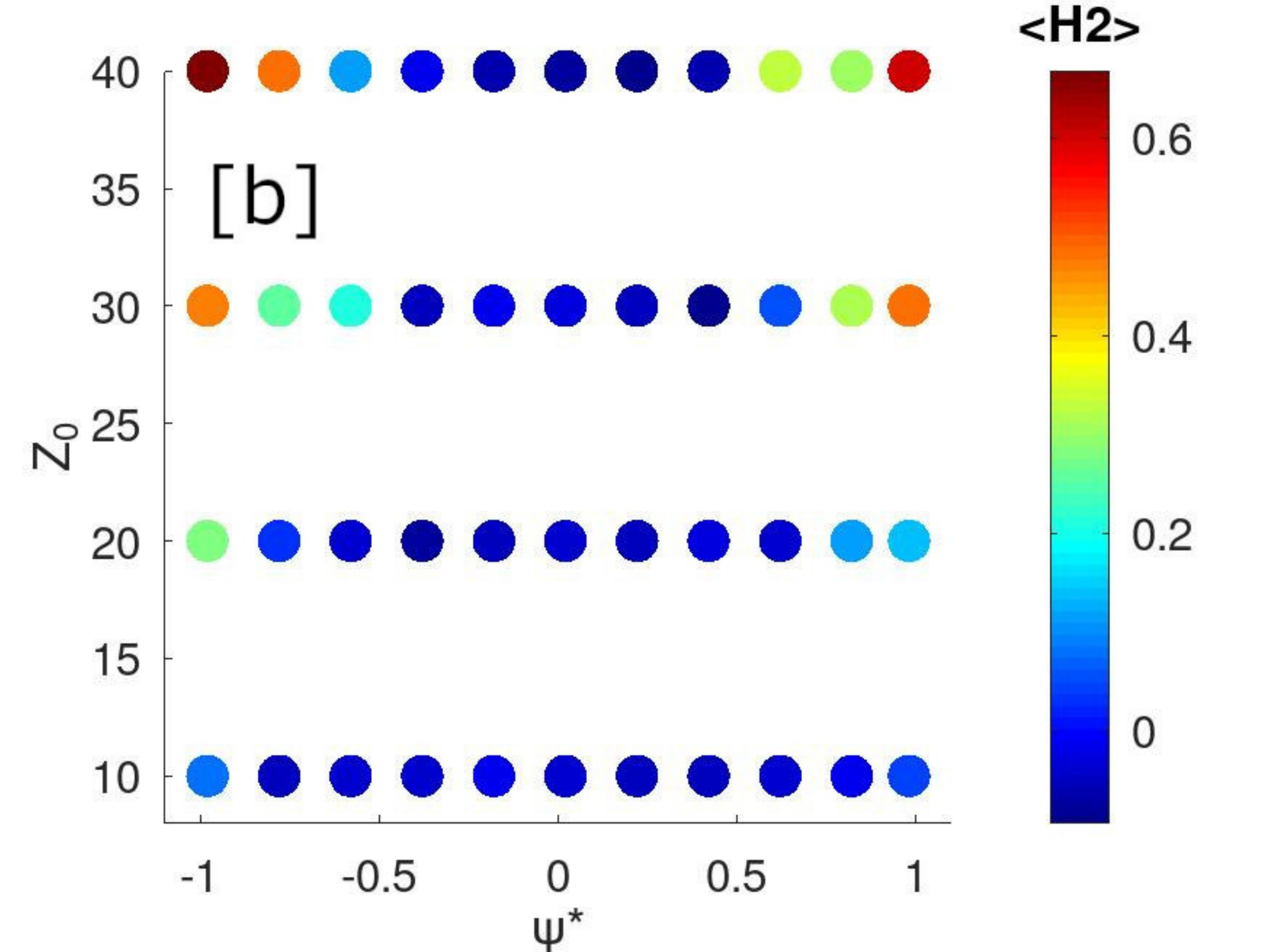}
\caption{\label{fig5}
{\bf State diagram of helix formation: }
Figure shows the parameter space conducive to helix formation. In (a) we show the phase diagram with respect 
to the Debye length $\lambda_d$ and the bending energy $\epsilon_B$. The maximum value of the charge in the polymer 
chain is $Z_0=40$ s.u and the asymmetry parameter $\psi=0.1$ or $\psi*=0.8$. In (b) we show the phase diagram 
with respect to the charge $Z_0$ and the asymmetry parameter $\psi*$ which is related to $\psi$ as 
$\psi*=-2(\psi-0.5)$. We keep the Debye length $\lambda_d=5a$ and the bending energy $\epsilon_B=200 k_BT$. 
In both (a) and (b) we switch on charges at $t=10\tau$ and the average is computed over an interval of $10\tau$.
}
\end{figure}

To quantify this unwinding process, we look at the periodicity parameter W as a function of time. W is essentially 
the dot product of bond vectors along the contour with a vector perpendicular to the axis of the helical polymer chain. We initialize the chain along the $\hat{y}$ direction. The helical axis of the emergent helix is also along $\hat{y}$.
Thus, we calculate W as 
\begin{equation}
  W (i)= \hat{x}.r_{i}
\end{equation}

where $r_i$ denotes the bond vectors along the chain contour, and $i$ denotes the monomer indices along the chain contour.
We now quantitatively establish that  the helical conformations of the polymer relax by gradual unwinding. 
To that end we plot in Fig.\ref{fig3} the periodicity parameter W at various times, during the simulation run. We note that the periodicity in W 
decreases as simulation time elapses.
 This indicates that the helical polymer gradually unwinds and the pitch keeps on increasing. At any given time there may be helical segments of varying pitch, which gradually increase. Note that the most frequently occurring pitch of the helix along the chain contour, can be found from the data presented in Fig.\ref{fig3} by conducting a Fourier transform \cite{Mitra2020}. 

 The most frequently occurring pitch of the helix $P$ at a particular time also depends on the value of the bending energy $\epsilon_B$. A high value of $\epsilon_B$ leads to larger values of $P$ at a particular instant of simulation time.  The other parameters do not affect the pitch of the helix.

\subsection{Parameters controlling helix formation}

There are $4$ system parameters which can be tuned to obtain the helical conformations in polymers. They are:
\begin{enumerate}
\item The magnitude of the charge on a monomer which is proportional to $Z_0$.  
\item The Debye length of the screened Coulomb interaction: $\lambda_D$.
\item The bending energy $\epsilon_B$, which is directly proportional to the persistence length \cite{Mitra2020}.
\item The degree of asymmetry in the charge distribution along the polymer chain.
\end{enumerate}  

 In Fig.\ref{fig4}, we investigate how helix formation depends on a balance of system parameters, 
 and thereby plot the time averaged $\langle H2\rangle$ as we vary the parameters one at a time. 
 Thereby, we keep the simulation parameters identical to those used to obtain data of  Fig.\ref{fig1},
 and vary one parameter at a time. We compute the time averaged $\langle H2\rangle$ over a duration of $10\tau$
 from $ t=10 \tau$ to $t=20\tau$ . 
 
 In Fig.\ref{fig4}a, we show that an increase in $Z_0$, i.e. maximum value and range of charges of the monomers 
 along the chain, leads to an increase in helical order obtained. We observe an increase in $\langle H2 \rangle$  with increase in $Z_0$ before 
 it plateaus around $0.6$. 
Moreover, the quantity $\langle H2 \rangle$  shows a  decrease with increase in $\epsilon_B$ as shown in  Fig.\ref{fig4}b.
 This is because an increase in the bending energy (or equivalently the persistence length $\ell_p$)
 prevents the formation of sharp angular  conformations as a response to the turning on of electrostatic forces. 
 As mentioned before, the formation of sharp angular 
 conformations (kinks) at time $ t=T_0$ are crucial to the subsequent formation of helices.
 Furthermore a higher chain rigidity also leads to the faster ``unwinding" and dissolution of the transient helices.
 
 An increase in the Debye length $\lambda_D$ leads to a larger number of electrostatic interactions between monomers and thus leads to a 
 greater propensity of helix formation. This leads to high values of $\langle H2\rangle$ for $\lambda_D$ in the 
 range of $5-10 a$.   Interestingly we see in Fig.\ref{fig4}c that $\langle H2\rangle$ shows a non-monotonic 
 dependence on $\lambda_D$. Note that here we have $\psi=0.1$ (or $\psi*=0.8$) which implies that the majority of 
 the charges along the chain have a  positive value. When $\lambda_D$ is increased then the strength of repulsive 
 interactions increase leading to an increase in the helicity. However when $\lambda_D$ is increased further then 
 there are also attractive interactions along the chain  (due to some monomers having charges of opposite polarity)
 that each monomer experiences. This leads to a decrease in the propensity for helix formation and consequently
 lower values for $\langle H2\rangle$ as there are strong attractions between different chain segments. 
 This in turn suppresses the effects of bending energy costs which led to helix formation. 
 We discuss the consequence of increasing $\lambda_D$ for a  lower value of $\psi$ subsequently.
 We emphasize that this suppression of helix formation is only due to the introduction of attractive interactions and not due to the increase in repulsive interactions. One may infer this from the data of Fig.\ref{fig4}a. There we show that increase in $Z_0$, leads to larger repulsive forces which increases the propensity of helix formation. 

 The asymmetry in the distribution of charges along the polymer chain plays a crucial role 
 in helix formation. In Fig.\ref{fig4}e we show how helicity as measured by $\langle H2\rangle$ depends 
 on the asymmetry parameter $\psi$ for a constant value of $Z_0$ which is set at $Z_0=40$s.u. For convenience, 
 we now define a new asymmetry parameter $\psi*$ which depends on $\psi$ as $\psi*=1 - 2\psi$. Note that $\psi*=+1$
 implies that all the charges along the polymer chain are negative while $\psi=-1$ denotes that all the charges are 
 positive. If $\psi*=0$ then there are statistically equal number of positive and negative charges 
 in the polymer chain. Thus greater the value of $|\psi*|$ greater is the asymmetry in the distribution of charges. 
 
 We see that for $\psi*\sim 0$ helicity goes down, which indicates that when the strength of attractive interactions 
 become comparable to the strength of repulsive interactions then helicity goes down. Thus helicity crucially relies 
 on the presence of strong repulsive interactions which may arise from both positive and negative charges. 
 This symmetry gives rise to the `U' shape seen in Fig.\ref{fig4}d.
 
 In Fig.\ref{fig4}a-c we examine the dependence of $\langle H2\rangle$ for a high value of asymmetry in the charge distribution ($\psi=0.1$ or $\psi*=0.8$). We also establish that the trends observed in Fig.\ref{fig4} are robust even for lower asymmetry in charge distribution, i.e for $\psi*=0.4$ (or $\psi=0.3$) albeit with a smaller value of $\langle H2\rangle$ as expected. This is shown in Fig.S1 of the Supplementary Information (SI).
 
Finally, we plot a state diagram using these parameters, using $\kappa=200 k_BT/a^2$ (same as in Fig.\ref{fig1})
to obtain the parameter space which is conducive to helix formation. In Fig.\ref{fig5}a we show the state diagram with 
Debye length $\lambda_D$ on the y-axis and the bending energy $\epsilon_B$ on the x-axis. We keep $Z_0$ and the charge distribution asymmetry 
parameter $\psi$ the same as in Fig.\ref{fig1}. We note that for $\lambda_D < 6a$, $\langle H2\rangle$ shows a
decrease with increase in $\epsilon_B$ beyond $1200k_BT$. This indicates that for relatively more rigid chains the helix 
formation is suppressed as has already  been shown in Fig.\ref{fig4}b. However, for intermediate values of $\lambda_D$ 
(i.e. $\lambda_D \approx 10a $), this suppression in propensity to form helices is not observed.  This is because
each monomer interacts electrostatically with a larger number of monomers along the polymer chain. However, for higher values 
of $\lambda_D$ ($\lambda_D > 15a$), the propensity to form helices again decreases. This is consistent with the non-monotonic behaviour of 
$\langle H2\rangle$ with increase in $\lambda_D$ as established in Fig.\ref{fig4}c and is for the same reason as explained. In Fig.\ref{fig5}b we plot another state diagram keeping the $\lambda_D$ and $\epsilon_B$ fixed with the values being the same as that of Fig.\ref{fig1}. Here we vary $Z_0$ and $\psi*$ to plot the state diagram. We note that for small values of $Z_0$, $\langle H2\rangle \approx 0$ while for higher values of $Z_0$, we obtain high values of  $\langle H2\rangle$ only when $\psi* \approx \pm 1$. These trends are consistent with those established in Fig.\ref{fig4}a and Fig.\ref{fig4}e. This once again emphasises that asymmetry in charge distribution of monomers along the chain is crucial to helix formation. \\

\begin{figure}[!hbt]
\includegraphics[width=0.48\columnwidth]{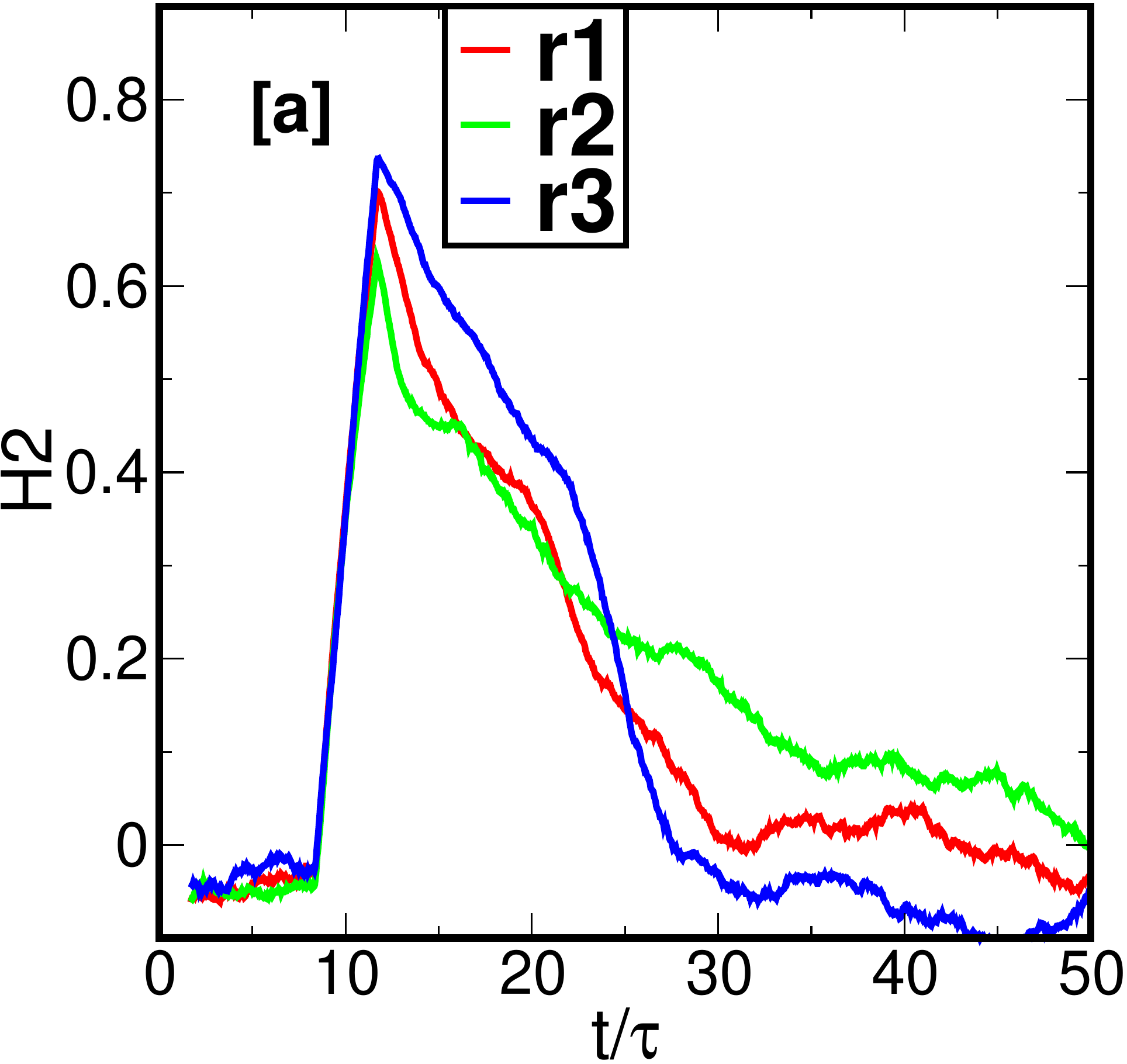}
\includegraphics[width=0.48\columnwidth]{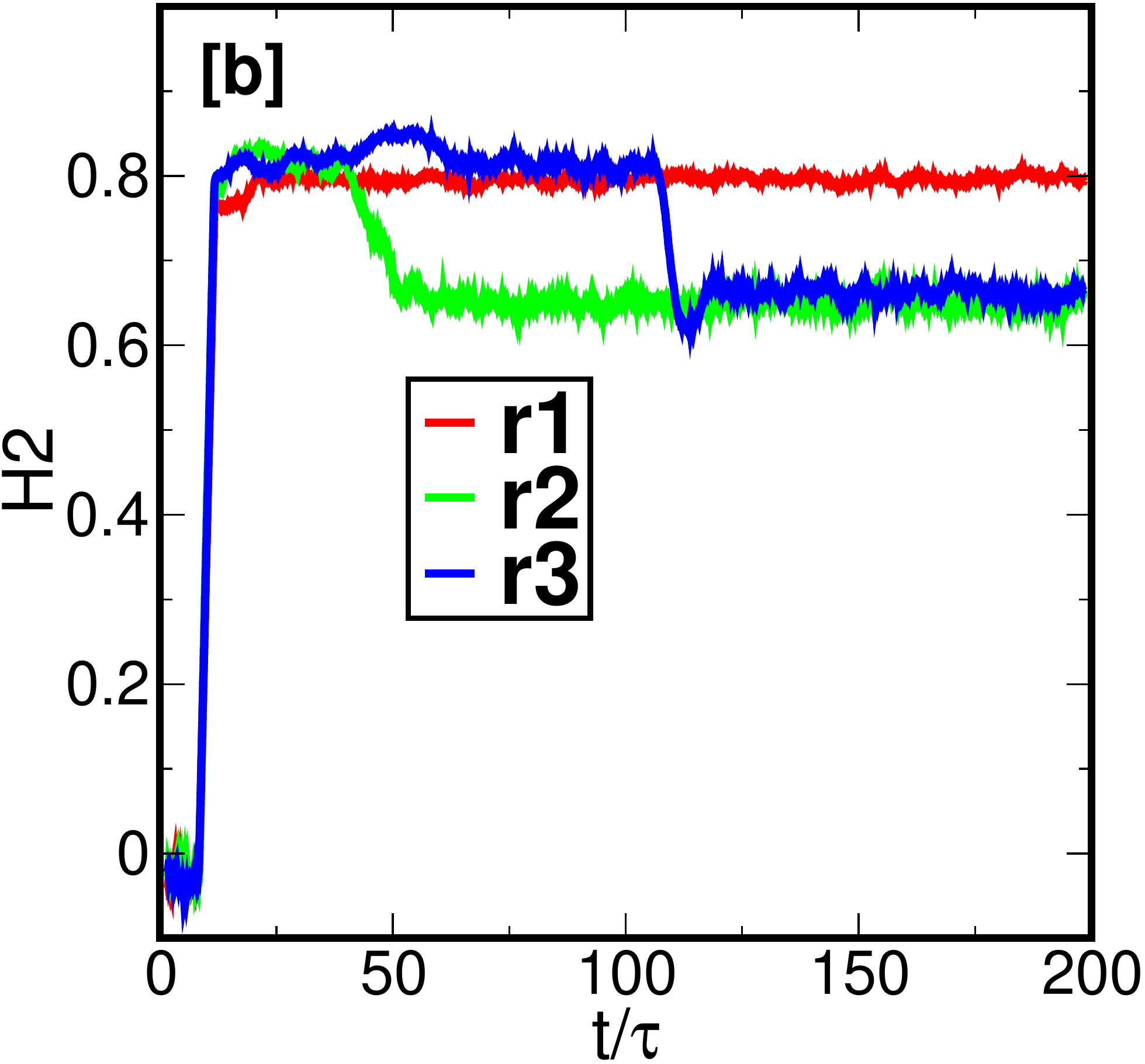}
\caption{\label{fig6}
{\bf Stabilization of transient helices:}
To examine whether geometric confinement is able to stabilize the transient helices we induce helix formation in a cuboid. The length of the cuboidal box $L_c = 1.1 R_c$, where $R_c$ is the contour length of the polymer chain having $50$ monomers. The width of the cuboid was  $W_c=0.1 R_c$. We implement walls of the cuboid by WCA interactions and we keep the parameters identical to that in Fig.\ref{fig1}. We switch on electrostatic interactions due to charges at $10 \tau$.  In (a) we show $H2$ (running averaged over $0.33\tau$) as a function of simulation time for an unconfined chain of $50$ monomers for $3$ independent runs. We see that the helix formed is transient and gradually dissolves with time. In (b) we show again a chain of $50$ monomers similar to that of (a) but under confinement. We see that $H2$ (running averaged over $0.33\tau$) shows a high value even at long times for $3$ independent runs. Thus geometric confinement lead to the stabilization of helices.}
\end{figure}

\subsection{Confinement stabilizes helices}
We now study whether imposing geometric confinement leads to the emergence of long-lived helices in our model. To investigate this, we simulate the same system confined within a cuboid, where the rigid walls 
which are modelled by inducing repulsive WCA interactions to prevent the monomers crossing the walls, refer Fig.\ref{fig6}. Thus we show 
that the transient helices formed due to electrostatic interactions between charged monomers may be stabilized or made long lived under confinement. 
In Fig.\ref{fig6}a we show that $H2$ or helicity for an unconfined chain dissolves away by $50\tau$ whereas the helicity ($H2$) for a 
(b) chain under confinement shows that $H2$ fluctuates about a high value even at long times. Thus geometric confinement leads to
stable helices due to electrostatic interactions.

How does geometric confinement lead to the formation of stable helices? As mentioned before, the transient helices dissolves as the pitch of the helix increases and the polymer chain stretches out axially. This is due to the predominantly repulsive, electrostatic interactions between monomers and the bending energy cost due to semiflexibility. However, if the polymer is placed in a cuboidal box with length $L_c = 1.1 R_c$  ($R_c$ is the initial contour length of the uncharged chain), the confinement prevents the stretching of the polymer chain axially and thus stops the unwinding process. This leads to the formation of long lived stable helices. Thus a geometric confinement with the axial dimension $L_c\sim R_c$ successfully leads to long lived helical structures. 
\begin{figure}[!hbt]
\includegraphics[width=1.2\columnwidth]{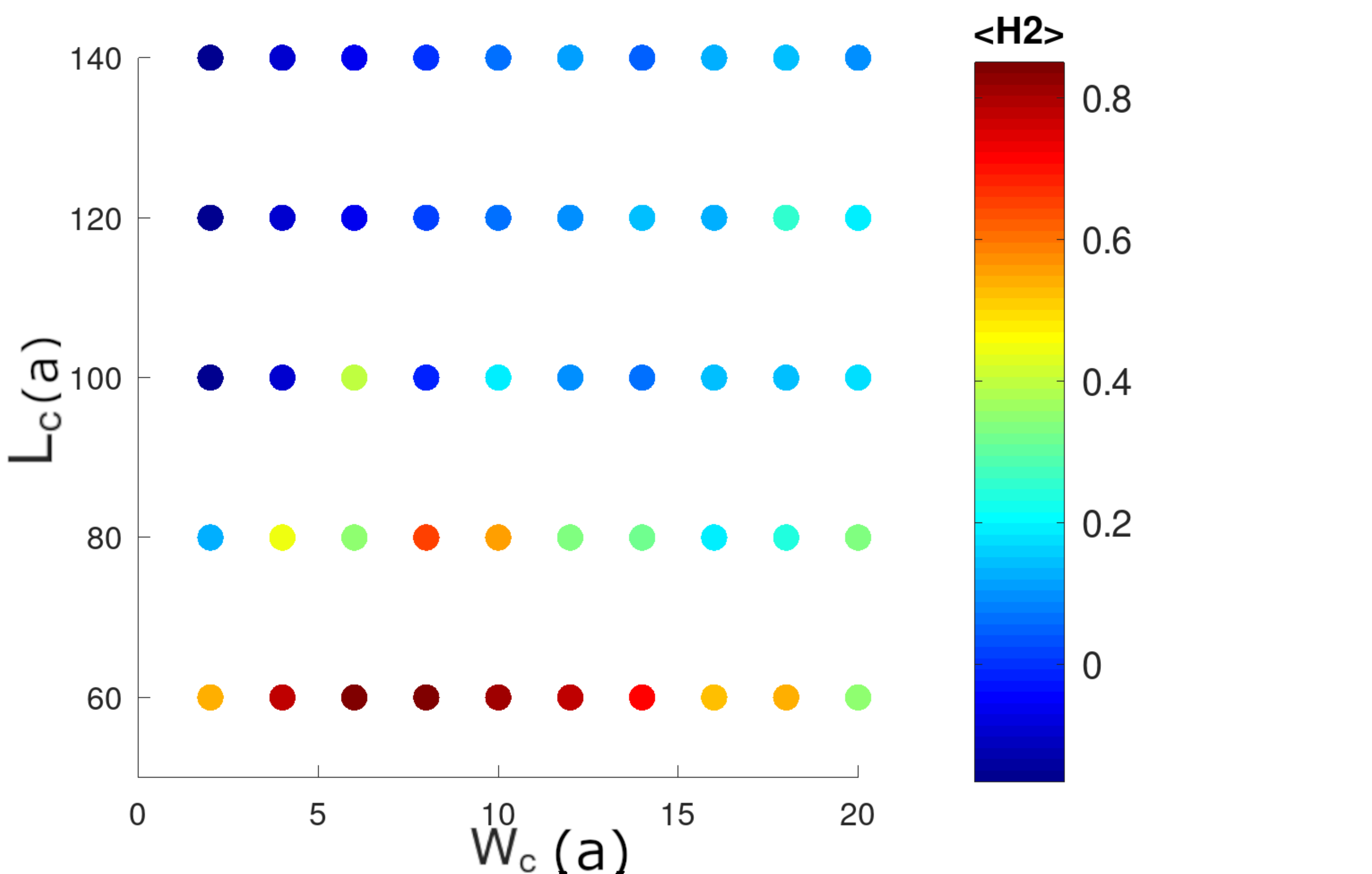}
\caption{\label{fig7}
{\bf Aspect ratios of cuboidal confinement conducive to helix formation:}
To examine the ranges of $W_c$ (width of the confining cuboid) and $L_c$  (length of the confining cuboid) for which we obtain stable helix formation. We switch on electrostatic interactions due to charges at $10 \tau$ and measure $\langle H2\rangle$ over the next $190\tau$. The other parameters are identical to that of Fig\ref{fig1}. The value of $R_c$ is $49 a$ where $a$ is the unit of length in our simulations.}
\end{figure}

This again leads to questions about the range of aspect ratios that are conducive 
to helix formation. To investigate this, we systematically vary the width $W_c$ and length of the cuboid $L_c$ of the cuboid with a square cross-section. We switch on charges at $t=10 \tau$ and compute $\langle H2\rangle$ over the next $190 \tau$. We plot a colour map showing the range of $W_c$ and $L_c$ conducive to stable helix formation, refer Fig.\ref{fig7}. We see that we obtain helix formation only when $L_c\sim R_c$ and also requires intermediate values of $W_c$. When the values of $W_c$ are small the monomers are unable to move significantly from the axis, thus leading to a suppression of helix formation. For large values of $W_c$ the polymer may stretch out diagonally as a response to repulsive forces. Thus one may attain long-lived, helical structures due to electrostatic interactions between charged monomers in a polymer under confinement with the aspect ratio chosen suitably. 

\section{Discussions}
Through this study, we have established a generic mechanism by which the helical motif may spontaneously emerge in several bio-molecules. We posit that such emergent structures may arise due to the induction of charges along the bio-polymer chain contour due to local changes in solvent conditions. However, these structures have to rely on other mechanisms to remain long-lived. Free-standing helices formed due to the switching on of screened Coulomb interactions are transient. In the presence of confinement, these helices become long-lived. In our study, we often use large Debye lengths ($\approx 10 a$). Such large values of the Debye length are unlikely to arise in natural polyelectrolytes. However, in this study we show that the mechanism of helix formation is robust even for such high values of the Debye lengths, which may be realized in experiments with highly deionized solvents.

The helices we observe in our simulations do not have a preferential handedness and may have an equal number of left-handed and right-handed segments. For most, bio-molecules however, there is a preferred handedness when the molecules attain helical conformations. Future investigations may shed light on this aspect and elucidate how a preferred handedness emerges. This will require incorporating specific constraints beyond the mechanism outlined here.

\section{Acknowledgements}
A.C., with DST-SERB identification SQUID-1973-AC-4067, acknowledges funding by DST-India, project MTR/2019/000078 and CRG/2021/007824. A.C also acknowledges discussions in meetings organized by ICTS, Bangalore.


\bibliographystyle{unsrt}
\bibliography{ref.bib}



\end{document}